\documentstyle[12pt,epsf]{article}

\newcommand{\agt}{\,\rlap{\lower 3.5 pt \hbox{$\mathchar \sim$}} \raise 1pt
 \hbox {$>$}\,}
\newcommand{\alt}{\,\rlap{\lower 3.5 pt \hbox{$\mathchar \sim$}} \raise 1pt
 \hbox {$<$}\,}
\newcommand{\re}{\mathop{\mbox{Re}}\nolimits}
\newcommand{\arcosh}{\mathop{\mbox{arcosh}}\nolimits}

\catcode`@=11
\newcount\@tempcntc
\def\@citex[#1]#2{\if@filesw\immediate\write\@auxout{\string\citation{#2}}\fi
  \@tempcnta\z@\@tempcntb\m@ne\def\@citea{}\@cite{\@for\@citeb:=#2\do
    {\@ifundefined
       {b@\@citeb}{\@citeo\@tempcntb\m@ne\@citea\def\@citea{,}{\bf ?}\@warning
       {Citation `\@citeb' on page \thepage \space undefined}}%
    {\setbox\z@\hbox{\global\@tempcntc0\csname b@\@citeb\endcsname\relax}%
     \ifnum\@tempcntc=\z@ \@citeo\@tempcntb\m@ne
       \@citea\def\@citea{,}\hbox{\csname b@\@citeb\endcsname}%
     \else
      \advance\@tempcntb\@ne
      \ifnum\@tempcntb=\@tempcntc
      \else\advance\@tempcntb\m@ne\@citeo
      \@tempcnta\@tempcntc\@tempcntb\@tempcntc\fi\fi}}\@citeo}{#1}}
\def\@citeo{\ifnum\@tempcnta>\@tempcntb\else\@citea\def\@citea{,}%
  \ifnum\@tempcnta=\@tempcntb\the\@tempcnta\else
   {\advance\@tempcnta\@ne\ifnum\@tempcnta=\@tempcntb \else \def\@citea{--}\fi
    \advance\@tempcnta\m@ne\the\@tempcnta\@citea\the\@tempcntb}\fi\fi}
\catcode`@=12

\begin{document}
\title{\vskip-3cm{\baselineskip14pt
\centerline{\normalsize DESY 03-096\hfill ISSN 0418-9833}
\centerline{\normalsize hep-ph/0307386\hfill}
\centerline{\normalsize July 2003\hfill}}
\vskip1.5cm
$\chi_{c1}$ and $\chi_{c2}$ decay angular distributions at the Fermilab
Tevatron}
\author{{\sc Bernd A. Kniehl, Gustav Kramer}\\
{\normalsize II. Institut f\"ur Theoretische Physik, Universit\"at Hamburg,}\\
{\normalsize Luruper Chaussee 149, 22761 Hamburg, Germany}\\
\\
{\sc Caesar P. Palisoc}\\
{\normalsize National Institute of Physics, University of the Philippines,}\\
{\normalsize Diliman, Quezon City 1101, Philippines}}

\date{}

\maketitle

\thispagestyle{empty}

\begin{abstract}
We consider the hadroproduction of $\chi_{c1}$ and $\chi_{c2}$ mesons and
their subsequent radiative decays to $J/\psi$ mesons and photons in the
factorization formalism of nonrelativistic quantum chromodynamics, and study
the decay angular distributions, by means of helicity density matrices, in
view of their sensitivity to color-octet processes.
We present numerical results appropriate for the Fermilab Tevatron.

\medskip

\end{abstract}

\newpage

\section{Introduction}

Since the discovery of the $J/\psi$ meson in 1974, charmonium has provided a
useful laboratory for quantitative tests of quantum chromodynamics (QCD) and,
in particular, of the interplay of perturbative and nonperturbative phenomena.
The factorization formalism of nonrelativistic QCD (NRQCD) \cite{bbl} provides
a rigorous theoretical framework for the description of heavy-quarkonium
production and decay.
This formalism implies a separation of short-distance coefficients, which can 
be calculated perturbatively as expansions in the strong-coupling constant
$\alpha_s$, from long-distance matrix elements (MEs), which must be extracted
from experiment.
The relative importance of the latter can be estimated by means of velocity
scaling rules, {\it i.e.}, the MEs are predicted to scale with a definite
power of the heavy-quark ($Q$) velocity $v$ in the limit $v\ll1$.
In this way, the theoretical predictions are organized as double expansions in
$\alpha_s$ and $v$.
A crucial feature of this formalism is that it takes into account the complete
structure of the $Q\overline{Q}$ Fock space, which is spanned by the states
$n={}^{2S+1}L_J^{(\zeta)}$ with definite spin $S$, orbital angular momentum
$L$, total angular momentum $J$, and color multiplicity $\zeta=1,8$.
In particular, this formalism predicts the existence of color-octet (CO)
processes in nature.
This means that $Q\overline{Q}$ pairs are produced at short distances in
CO states and subsequently evolve into physical, color-singlet (CS) quarkonia
by the nonperturbative emission of soft gluons.
In the limit $v\to0$, the traditional CS model (CSM) \cite{ber} is recovered.
The greatest triumph of this formalism was that it was able to correctly 
describe \cite{bra} the cross section of inclusive charmonium
hadroproduction measured in $p\overline{p}$ collisions at the Fermilab
Tevatron \cite{abe}, which had turned out to be more than one order of
magnitude in excess of the theoretical prediction based on the CSM.

Apart from this phenomenological drawback, the CSM also suffers from severe
conceptual problems indicating that it is incomplete.
These include the presence of logarithmic infrared divergences in the
${\cal O}(\alpha_s)$ corrections to $P$-wave decays to light hadrons and in
the relativistic corrections to $S$-wave annihilation \cite{bar}, and the lack
of a general argument for its validity in higher orders of perturbation
theory.
While the $k_T$-factorization \cite{sri} and hard-comover-scattering
\cite{hoy} approaches manage to bring the CSM prediction much closer to the
Tevatron data, they do not cure the conceptual defects of the CSM.
The color evaporation model \cite{cem}, which is intuitive and useful for
qualitative studies, also leads to a significantly better description of the
Tevatron data, but it is not meant to represent a rigorous framework for
perturbation theory.
In this sense, a coequal alternative to the NRQCD factorization formalism is 
presently not available.

In order to convincingly establish the phenomenological significance of the
CO processes, it is indispensable to identify them in other kinds of
high-energy experiments as well.
Studies of charmonium production in $ep$ photoproduction, $ep$ and $\nu N$
deep-inelastic scattering (DIS), $e^+e^-$ annihilation, $\gamma\gamma$
collisions, and $b$-hadron decays may be found in the literature; see
Ref.~\cite{fle} and references cited therein.
Furthermore, the polarization of $\psi^\prime$ mesons produced directly and of
$J/\psi$ mesons produced promptly, {\it i.e.}, either directly or via the 
feed-down from heavier charmonia, which also provides a sensitive probe of CO
processes, was investigated \cite{ben,van,bkl,lee}.
Until very recently, none of these studies was able to prove or disprove the
NRQCD factorization hypothesis.
However, H1 data of $ep\to eJ/\psi+X$ in DIS at HERA \cite{h1} and DELPHI data
of $\gamma\gamma\to J/\psi+X$ at LEP2 \cite{delphi} provide first
independent evidence for it \cite{dis,gg}.

The cross section ratio of $\chi_{c1}$ and $\chi_{c2}$ inclusive
hadroproduction was measured by the CDF Collaboration in Run~1 at the Tevatron
\cite{aff}.
The $\chi_{cJ}$ mesons were detected through their radiative decays to
$J/\psi$ mesons and photons.
The $J/\psi$ mesons were identified via their decay to $\mu^+\mu^-$ pairs,
and the photons were reconstructed through their conversion to $e^+e^-$ pairs,
which provide an excellent energy resolution for the primary photons.
The increased data sample to be collected during Run~2, which has just 
started, will allow for more detailed investigations of these cross sections,
including the angular distributions of the $\chi_{cJ}$ decay products.
The latter carry all the information on the $\chi_{cJ}$ polarization and thus
provide a handle on the CO processes.

In this paper, we study the angular distributions of the processes
$p\overline{p}\to\chi_{cJ}+X$ ($J=1,2$) with subsequent decays
$\chi_{cJ}\to J/\psi\gamma$ with regard to their power to distinguish between
NRQCD and the CSM.
Specifically, we consider the polar and azimuthal angles, $\theta$ and $\phi$,
of the $J/\psi$ meson in the $\chi_{cJ}$ rest frame.
Once a suitable coordinate system is defined in that frame, the full angular
information is encoded in the helicity density matrix
$\rho_{\lambda\lambda^\prime}^J$ of the primary $\chi_{cJ}$ production
process, where $\lambda$ and $\lambda^\prime$ denote the $\chi_{cJ}$
helicities.
The matrix $\rho_{\lambda\lambda^\prime}^J$, therefore, provides a systematic
way of analyzing the complicated details of the $\theta$ and $\phi$
dependences.
In the following, we study $\rho_{\lambda\lambda^\prime}^J$ in four commonly
used polarization frames as functions of the $\chi_{cJ}$ transverse momentum
$p_T$ so as to identify optimal observables to discriminate NRQCD from the
CSM.

This paper is organized as follows.
In Sec.~\ref{sec:ana}, we list the formulas necessary to evaluate the helicity
density matrix of $p\overline{p}\to\chi_{cJ}+X\to J/\psi\gamma+X$ for $J=1,2$.
In Sec.~\ref{sec:num}, we present our numerical results and discuss their
phenomenological implications.
Our conclusions are summarized in Sec.~\ref{sec:con}.

\section{Analytic results}
\label{sec:ana}

We consider the production processes $p\overline{p}\to\chi_{cJ}j+X$, where
$J=1,2$ and $j$ denotes a hadron jet, followed by the radiative decays
$\chi_{cJ}\to J/\psi\gamma$ in the narrow-width approximation.
(The presence of the hadron jet, which does not have to be observed, allows
for the $\chi_{cJ}$ meson to have finite transverse momentum.)
Let $\sqrt S$ be the center-of-mass energy of the hadronic collision, $M$ the
mass of the $\chi_{cJ}$ meson, $p_T$ the transverse momentum common to the
$\chi_{cJ}$ meson and the jet, $y$ and $y_j$ the rapidities of the latter,
$\lambda=-J,\ldots,J$ the helicity of the $\chi_{cJ}$ meson in some
polarization frame, and $\theta$ and $\phi$ the polar and azimuthal angles of
the $J/\psi$ meson in the respective coordinate system defined in the
$\chi_{cJ}$ rest frame.
Then, the differential cross section can be written as
\begin{equation}
\frac{d^5\sigma^J}{dp_T^2\,dy\,dy_j\,d^2\Omega}
=B_J\sum_{\lambda,\lambda^\prime=-J}^J
\frac{d^3\rho_{\lambda\lambda^\prime}^J}{dp_T^2\,dy\,dy_j}
A_{\lambda\lambda^\prime}^J(\theta,\phi),
\label{eq:xs}
\end{equation}
where $\rho_{\lambda\lambda^\prime}^J$ and $A_{\lambda\lambda^\prime}^J$ refer
to the production and decay processes, respectively, $B_J$ denotes the
branching fraction of $\chi_{cJ}\to J/\psi\gamma$, and
$d^2\Omega=d\cos\theta\,d\phi$.
Invoking the factorization theorems of the QCD parton model and NRQCD, we have
\begin{equation}
\frac{d^3\rho_{\lambda\lambda^\prime}^J}{dp_T^2\,dy\,dy_j}
=\sum_{a,b,d,n}x_af_{a/p}(x_a,\mu_f)x_bf_{b/\overline{p}}(x_b,\mu_f)
\langle{\cal O}^{\chi_{cJ}}[n]\rangle\frac{1}{16\pi s}
\rho_{\lambda\lambda^\prime}\left(ab\to c\overline{c}[n]d\right).
\end{equation}
Here, it is summed over the active partons $a,b,d=g,q,\overline{q}$ and the
$c\overline{c}$ Fock states $n$,
$f_{a/h}(x_a,\mu_f)$ are the PDFs of the beam hadron $h$,
$x_a$ is the fraction of longitudinal momentum that $a$ receives from $h$,
$\mu_f$ is the factorization scale,
$\langle{\cal O}^{\chi_{cJ}}[n]\rangle$ are the MEs of the $\chi_{cJ}$ meson,
and
\begin{equation}
\rho_{\lambda\lambda^\prime}\left(ab\to c\overline{c}[n]d\right)
=\sum{\cal T}^\star\left(ab\to c\overline{c}[n,\lambda^\prime]d\right)
{\cal T}\left(ab\to c\overline{c}[n,\lambda]d\right),
\label{eq:rho}
\end{equation}
where it is averaged (summed) over the spin and color states of $a$ and $b$
($d$), and ${\cal T}$ denotes the transition-matrix element.
We have
\begin{equation}
x_{a,b}=\frac{m_T\exp(\pm y)+p_T\exp(\pm y_j)}{\sqrt S},
\end{equation}
where $m_T=\sqrt{M^2+p_T^2}$ is the transverse mass of the $\chi_{cJ}$ meson.
The partonic Mandelstam variables $s=(p_a+p_b)^2$, $t=(p_a-P)^2$, and
$u=(p_b-P)^2$, where $P$ is the four-momentum of the $\chi_{cJ}$ meson, can be
expressed in terms of $p_T$, $y$, and $y_j$ as
\begin{eqnarray}
s&=&p_T^2+m_T^2+2p_Tm_T\cosh(y-y_j),
\nonumber\\
t&=&-p_T^2-p_Tm_T\exp(y_j-y),
\nonumber\\
u&=&-p_T^2-p_Tm_T\exp(y-y_j),
\end{eqnarray}
respectively.
Notice that $s+t+u=M^2$ and $sp_T^2=tu$.
The kinematically allowed ranges of $p_T$, $y$, and $y_j$ are
\begin{eqnarray}
0&\le&p_T\le\frac{S-M^2}{2\sqrt{S}},
\nonumber\\
|y|&\le&\arcosh\frac{S+M^2}{2\sqrt Sm_T},
\nonumber\\
-\ln\frac{\sqrt S-m_T\exp(-y)}{p_T}&\le&y_j\le
\ln\frac{\sqrt S-m_T\exp(y)}{p_T}.
\end{eqnarray}
Integrating Eq.~(\ref{eq:xs}) over $y_j$, we obtain the inclusive cross
section of $p\overline{p}\to\chi_{cJ}+X$.

Choosing a suitable coordinate system in the $\chi_{cJ}$ rest frame and
defining
\begin{eqnarray}
\langle\lambda_1,\lambda_2,\theta,\phi|T|J,\lambda\rangle
&=&{\cal T}(\chi_{cJ}(\lambda)\to J/\psi(\lambda_1,\theta,\phi)
\gamma(\lambda_2,\pi-\theta,\phi+\pi)),
\nonumber\\
T_{\lambda_1\lambda_2}^J&=&\sqrt{\frac{4\pi}{2J+1}}
\left.\langle\lambda_1,\lambda_2,0,0|T|J,\lambda\rangle\right|_{\lambda=
\lambda_1-\lambda_2},
\end{eqnarray}
where $\lambda_1=0,\pm1$ and $\lambda_2=\pm1$ are the helicities of the
$J/\psi$ and $\gamma$ bosons, respectively, the decay angular distribution is
encoded in the matrix
\begin{equation}
A_{\lambda\lambda^\prime}^J(\theta,\phi)
=\frac{\sum_{\lambda_1,\lambda_2}
\langle\lambda_1,\lambda_2,\theta,\phi|T|J,\lambda^\prime\rangle^\star
\langle\lambda_1,\lambda_2,\theta,\phi|T|J,\lambda\rangle}
{\sum_{\lambda_1,\lambda_2}|T_{\lambda_1\lambda_2}^J|^2}.
\end{equation}
This expression may be conveniently evaluated by observing that \cite{pil}
\begin{equation}
\langle\lambda_1,\lambda_2,\theta,\phi|T|J,\lambda\rangle
=\sqrt{\frac{2J+1}{4\pi}}
D_{\lambda,\lambda_1-\lambda_2}^{J\star}(-\phi,\theta,\phi)
T_{\lambda_1\lambda_2}^J,
\end{equation}
where
$D_{m^\prime m}^j(\alpha,\beta,\gamma)
=\langle j,m^\prime|D(\alpha,\beta,\gamma)|j,m\rangle$
is the representation of the rotation operator
$D(\alpha,\beta,\gamma)=\exp(-i\gamma J_z)\exp(-i\beta J_y)
\exp(-i\alpha J_z)$,
with $\alpha$, $\beta$, and $\gamma$ being the Euler angles, in the
eigenstates $|j,m\rangle$ of ${\boldmath J}^2$ and $J_z$.
We have
\begin{equation}
D_{m^\prime m}^j(\alpha,\beta,\gamma)
=\exp(-i\gamma m^\prime)d_{m^\prime m}^j(\beta)\exp(-i\alpha m),
\end{equation}
where
$d_{m^\prime m}^j(\beta)=\langle j,m^\prime|\exp(-i\beta J_y)|j,m\rangle$
are the well-known $d$ functions, which may be evaluated from Wigner's formula
\cite{sak}
\begin{eqnarray}
d_{m^\prime m}^j(\beta)
&=&\sum_{k=\max(0,m-m^\prime)}^{\min(j+m,j-m^\prime)}(-1)^{k-m+m^\prime}
\frac{\sqrt{(j+m)!(j-m)!(j+m^\prime)!(j-m^\prime)!}}
{k!(k-m+m^\prime)!(j+m-k)!(j-m^\prime-k)!}
\nonumber\\
&&{}\times
\left(\cos\frac{\beta}{2}\right)^{2j+m-m^\prime-2k}
\left(\sin\frac{\beta}{2}\right)^{2k-m+m^\prime}.
\end{eqnarray}
Owing to the orthogonality relation
\begin{equation}
\int d^2\Omega
D_{m^\prime m^{\prime\prime}}^{j}(-\phi,\theta,\phi)
D_{m m^{\prime\prime}}^{j\star}(-\phi,\theta,\phi)
=\frac{4\pi}{2j+1}\delta_{m^\prime m},
\end{equation}
$A_{\lambda\lambda^\prime}^J$ is normalized as
\begin{equation}
\int d^2\Omega A_{\lambda\lambda^\prime}^J(\theta,\phi)
=\delta_{\lambda\lambda^\prime},
\end{equation}
so that, upon integration over the solid angle, Eq.~(\ref{eq:xs}) reduces to
the unpolarized narrow-width approximation formula
\begin{equation}
\frac{d^3\sigma}{dp_T^2\,dy\,dy_j}
\left(p\overline{p}\to\chi_{cJ}(\to J/\psi\gamma)j+X\right)
=B(\chi_{cJ}\to J/\psi\gamma)\frac{d^3\sigma}{dp_T^2\,dy\,dy_j}
\left(p\overline{p}\to\chi_{cJ}j+X\right).
\end{equation}

By definition, the helicity density matrix $\rho_{\lambda\lambda^\prime}^J$ is
hermitian,
$\rho_{\lambda^\prime\lambda}^{J\star}=\rho_{\lambda\lambda^\prime}^J$.
Furthermore, invariance of the ${\cal T}$ matrix under reflection in the
production plane, by action of the operator
${\cal Y}=\exp(-i\pi J_y){\cal P}$, where ${\cal P}$ is the parity operator,
leads to the symmetry property
$\rho_{-\lambda,-\lambda^\prime}^J
=(-1)^{\lambda-\lambda^\prime}\rho_{\lambda\lambda^\prime}^J$ \cite{pil}.
Thanks to these constraints, the number of independent real quantities 
contained in $\rho_{\lambda\lambda^\prime}^J$ is 5 for $J=1$ and 13 for $J=2$.
We may select the entries $\rho_{\lambda\lambda^\prime}^J$ with
$\lambda=0,\dots,J$ and $\lambda^\prime=-\lambda,\ldots,\lambda$, noticing
that they are real if $|\lambda^\prime|=\lambda$.

As for the decay amplitude $T_{\lambda_1\lambda_2}^J$, angular-momentum
conservation imposes the selection rule $|\lambda_1-\lambda_2|\le J$.
Furthermore, parity conservation entails the symmetry property \cite{pil}
\begin{eqnarray}
T_{-\lambda_1,-\lambda_2}^J
&=&\eta\eta_1\eta_2(-1)^{J_1+J_2-J}T_{\lambda_1\lambda_2}^J
\nonumber\\
&=&(-1)^JT_{\lambda_1\lambda_2}^J,
\end{eqnarray}
where $\eta$, $\eta_1$, and $\eta_2$ ($J$, $J_1$, and $J_2$) are the parity
(total-angular-momentum) quantum numbers of the $\chi_{cJ}$, $J/\psi$, and
$\gamma$ bosons, respectively.
An independent set of $T_{\lambda_1\lambda_2}^J$ amplitudes thus reads
\begin{eqnarray}
t_0^1&=&T_{1,1}^1=-T_{-1,-1}^1,\nonumber\\
t_1^1&=&T_{0,-1}^1=-T_{0,1}^1,\nonumber\\
t_0^2&=&T_{1,1}^2=T_{-1,-1}^2,\nonumber\\
t_1^2&=&T_{0,-1}^2=T_{0,1}^2,\nonumber\\
t_2^2&=&T_{1,-1}^2=T_{-1,1}^2.
\end{eqnarray}

Using these ingredients, we can now work out the general form of the angular
distributions of the $J/\psi$ meson,
\begin{equation}
W^J(\theta,\phi)=\sum_{\lambda,\lambda^\prime=-J}^J
\rho_{\lambda\lambda^\prime}^JA_{\lambda\lambda^\prime}^J(\theta,\phi).
\end{equation}
We find
\begin{eqnarray}
W^1(\theta,\phi)&=&\frac{3}{4\pi}\left\{
\rho_{0,0}^1\left[r_0^1\cos^2\theta+\frac{r_1^1}{2}\sin^2\theta\right]
+\rho_{1,1}^1\left[r_0^1\sin^2\theta+\frac{r_1^1}{2}(1+\cos^2\theta)\right]
\right.
\nonumber\\
&&{}-\left.
\sqrt2\re\rho_{1,0}^1\left(2r_0^1-r_1^1\right)\sin\theta\cos\theta\cos\phi
-\rho_{1,-1}^1\left(r_0^1-\frac{r_1^1}{2}\right)\sin^2\theta\cos(2\phi)
\right\},
\nonumber\\
W^2(\theta,\phi)&=&\frac{5}{4\pi}\left\{
\frac{\rho_{0,0}^2}{2}\left[\frac{r_0^2}{2}(1-3\cos^2\theta)^2
+3r_1^2\sin^2\theta\cos^2\theta+\frac{3}{4}r_2^2\sin^4\theta\right]
\right.
\nonumber\\
&&{}+\rho_{1,1}^2\left[3r_0^2\sin^2\theta\cos^2\theta
+\frac{r_1^2}{2}(1-3\cos^2\theta+4\cos^4\theta)
+\frac{r_2^2}{2}\sin^2\theta(1+\cos^2\theta)\right]
\nonumber\\
&&{}+\rho_{2,2}^2\left[\frac{3}{4}r_0^2\sin^4\theta
+\frac{r_1^2}{2}\sin^2\theta(1+\cos^2\theta)
+\frac{r_2^2}{8}(1+6\cos^2\theta+\cos^4\theta)\right]
\nonumber\\
&&{}+\sqrt6\re\rho_{1,0}^2
\left[r_0^2(1-3\cos^2\theta)
-r_1^2(1-2\cos^2\theta)
+\frac{r_2^2}{2}\sin^2\theta\right]
\sin\theta\cos\theta\cos\phi
\nonumber\\
&&{}-\re\rho_{2,1}^2
\left[3r_0^2\sin^2\theta
+2r_1^2\cos^2\theta
-\frac{r_2^2}{2}(3+\cos^2\theta)\right]
\sin\theta\cos\theta\cos\phi
\nonumber\\
&&{}-\rho_{1,-1}^2
\left[3r_0^2\cos^2\theta
+\frac{r_1^2}{2}(1-4\cos^2)
-\frac{r_2^2}{2}\sin^2\theta\right]
\sin^2\theta\cos(2\phi)
\nonumber\\
&&{}-\sqrt6\re\rho_{2,0}^2
\left[\frac{r_0^2}{2}(1-3\cos^2\theta)
+r_1^2\cos^2\theta
-\frac{r_2^2}{4}(1+\cos^2\theta)\right]
\sin^2\theta\cos(2\phi)
\nonumber\\
&&{}+\re\rho_{2,-1}^2\left(3r_0^2-2r_1^2+\frac{r_2^2}{2}\right)
\sin^3\theta\cos\theta\cos(3\phi)
\nonumber\\
&&{}+\left.\frac{\rho_{2,-2}^2}{2}\left(\frac{3}{2}r_0^2-r_1^2
+\frac{r_2^2}{4}\right)
\sin^4\theta\cos(4\phi)
\right\},
\label{eq:w}
\end{eqnarray}
where
\begin{equation}
r_\lambda^J=\frac{\left|t_\lambda^J\right|^2}
{\sum_{\lambda^\prime=0}^J\left|t_{\lambda^\prime}^J\right|^2}
\end{equation}
are positive numbers satisfying $\sum_{\lambda=0}^Jr_\lambda^J=1$ to be
determined experimentally.
In fact, the Fermilab E835 Collaboration \cite{e835} studied the angular
distributions of the exclusive reactions
$p\overline{p}\to\chi_{cJ}\to J/\psi\gamma\to e^+e^-\gamma$, with $J=1,2$,
and measured the fractional amplitudes of the electric dipole (E1), magnetic
quadrupole (M2), and electric octupole (E3) transitions.
They found the E1 transition to dominate, as expected from theoretical 
considerations.
From their results, we extract the values
\begin{eqnarray}
r_0^1&=&0.498\pm0.032,\qquad
r_1^1=0.502\pm0.032,
\nonumber\\
r_0^2&=&0.075\pm0.029,\qquad
r_1^2=0.250\pm0.048,\qquad
r_2^2=0.674\pm0.052.
\end{eqnarray}
The corresponding values for a pure E1 transition read
$r_0^1=r_1^1=0.5$, $r_0^2=0.1$, $r_1^2=0.3$, and $r_2^2=0.6$.

We work in the fixed-flavor-number scheme, {\it i.e.}, we have $n_f=3$ active
quark flavors $q=u,d,s$ in the proton and antiproton.
As required by parton-model kinematics, we treat the quarks $q$ as massless.
The charm quark $c$ and antiquark $\overline{c}$ only appear in the final 
state.
We are thus led to consider the following partonic subprocesses:
\begin{eqnarray}
gg&\to&c\overline{c}[n]g,
\nonumber\\
gq&\to&c\overline{c}[n]q,
\nonumber\\
q\overline{q}&\to&c\overline{c}[n]g.
\label{eq:pro}
\end{eqnarray}
As for the $\chi_{cJ}$ mesons, the $c\overline{c}$ Fock states contributing at
LO in $v$ are $n={}^3\!P_J^{(1)},{}^3\!S_1^{(8)}$.
Their MEs satisfy the multiplicity relations
\begin{eqnarray}
\left\langle{\cal O}^{\chi_{cJ}}\left[{}^3\!P_J^{(1)}\right]\right\rangle
&=&(2J+1)
\left\langle{\cal O}^{\chi_{c0}}\left[{}^3\!P_0^{(1)}\right]\right\rangle,
\nonumber\\
\left\langle{\cal O}^{\chi_{cJ}}\left[{}^3\!S_1^{(8)}\right]\right\rangle
&=&(2J+1)
\left\langle{\cal O}^{\chi_{c0}}\left[{}^3\!S_1^{(8)}\right]\right\rangle,
\label{eq:mul}
\end{eqnarray}
which follow to LO in $v$ from heavy-quark spin symmetry.

Depending on the value of $J$, the partonic helicity density matrices defined
in Eq.~(\ref{eq:rho}) may be decomposed as
\begin{eqnarray}
\rho_{\lambda\lambda^\prime}\left(ab\to c\overline{c}
\left[{}^{2S+1}L_1^{(\zeta)}\right]d\right)
&=&\epsilon_\mu^\star(\lambda)\epsilon_\nu(\lambda^\prime)
F\left[c_1g^{\mu\nu}
+c_2p_a^\mu p_a^\nu
+c_3p_b^\mu p_b^\nu
+\frac{c_4}{2}\left(p_a^\mu p_b^\nu+p_a^\nu p_b^\mu\right)\right],
\nonumber\\
\rho_{\lambda\lambda^\prime}\left(ab\to c\overline{c}
\left[{}^{2S+1}L_2^{(\zeta)}\right]d\right)
&=&\epsilon_{\mu\nu}^\star(\lambda)\epsilon_{\rho\sigma}(\lambda^\prime)
F\left[c_1g^{\mu\rho}g^{\nu\sigma}
+c_2g^{\mu\rho}p_a^\nu p_a^\sigma
+c_3g^{\mu\rho}p_b^\nu p_b^\sigma
\vphantom{\frac{c_9}{2}}
\right.
\nonumber\\
&&{}
+\frac{c_4}{2}g^{\mu\rho}\left(p_a^\nu p_b^\sigma+p_a^\sigma p_b^\nu\right)
+c_5p_a^\mu p_a^\nu p_a^\rho p_a^\sigma
+c_6p_b^\mu p_b^\nu p_b^\rho p_b^\sigma
\nonumber\\
&&{}
+\frac{c_7}{2}p_a^\mu p_a^\rho\left(p_a^\nu p_b^\sigma+p_a^\sigma p_b^\nu
\right)
+\frac{c_8}{2}p_b^\mu p_b^\rho\left(p_b^\nu p_a^\sigma+p_b^\sigma p_a^\nu
\right)
\nonumber\\
&&{}+\left.
\frac{c_9}{2}\left(p_a^\mu p_a^\nu p_b^\rho p_b^\sigma
+p_a^\rho p_a^\sigma p_b^\mu p_b^\nu\right)
+c_{10}p_a^\mu p_b^\nu p_a^\rho p_b^\sigma\right],
\end{eqnarray}
where $\epsilon^\mu(\lambda)$ ($\epsilon^{\mu\nu}(\lambda)$) is the
polarization vector (tensor) of a $J=1$ ($J=2$) boson with mass $M$,
four-momentum $P$, and helicity $\lambda$, and $F$ and $c_i$ are functions of
the partonic mandelstam variables $s$, $t$, and $u$.
Since we work at the tree level, $F$ and $c_i$ are real, so that the imaginary
parts of $\rho_{\lambda\lambda^\prime}^J$ all arise from
$\epsilon^\mu(\lambda)$ and $\epsilon^{\mu\nu}(\lambda)$.
The functions $F$ and $c_i$ for processes~(\ref{eq:pro}) with
$n={}^3\!P_1^{(1)},{}^3\!P_2^{(1)}$ are listed in the Appendix of
Ref.~\cite{lee}.
The results for $n={}^3\!S_1^{(8)}$ may be found in Eqs.~(B27)--(B31),
(B39)--(B43), and (B51)--(B55) of Ref.~\cite{van}, where one has to identify
$c_1=a$, $c_2=M^2b$, $c_3=M^2c$, and $c_4=2M^2d$.

Lorentz-covariant expressions for $\epsilon^\mu(\lambda)$ in four commonly
used polarization frames, namely, the recoil, Gottfried-Jackson, target, and
Collins-Soper frames, may be found in Ref.~\cite{lam}.
These frames differ in the way $\epsilon^\mu(0)$ is fixed.
In the $\chi_{cJ}$ rest frame, where
$\epsilon^\mu(0)=(0,\mbox{\boldmath$\epsilon$}(0))$, we have
$\mbox{\boldmath$\epsilon$}(0)=-\mbox{\boldmath$\hat p$}$, where
$p=p_p+p_{\overline{p}}$, in the recoil frame;
$\mbox{\boldmath$\epsilon$}(0)=\mbox{\boldmath$\hat p_p$}$ in the
Gottfried-Jackson frame;
$\mbox{\boldmath$\epsilon$}(0)=-\mbox{\boldmath$\hat p_{\overline{p}}$}$ 
in the target frame; and
$\mbox{\boldmath$\epsilon$}(0)=\left(\mbox{\boldmath$\hat p_p$}
-\mbox{\boldmath$\hat p_{\overline{p}}$}\right)/
\sqrt{2\left(1-\mbox{\boldmath$\hat p_p$}\cdot
\mbox{\boldmath$\hat p_{\overline{p}}$}\right)}$
in the Collins-Soper frame.
Here,
$\mbox{\boldmath$\hat p$}=\mbox{\boldmath$p$}/|\mbox{\boldmath$p$}|$ denotes
the unit three-vector.
With the help of the addition theorem for two angular momenta,
$\epsilon^{\mu\nu}(\lambda)$ can be constructed from the polarization
four-vectors of two $J=1$ bosons with the same four-momentum as \cite{gub}
\begin{equation}
\epsilon^{\mu\nu}(\lambda)=\sum_{\lambda_1,\lambda_2=-1}^1
\langle1,\lambda_1;1,\lambda_2|2,\lambda\rangle
\epsilon^\mu(\lambda_1)\epsilon^\nu(\lambda_2),
\end{equation}
where $\langle1,\lambda_1;1,\lambda_2|2,\lambda\rangle$ are Clebsch-Gordon
coefficients.
We have $P_\mu\epsilon^\mu(\lambda)=P_\mu\epsilon^{\mu\nu}(\lambda)=
g_{\mu\nu}\epsilon^{\mu\nu}(\lambda)=0$ and
$\epsilon^{\mu\nu}(\lambda)=\epsilon^{\nu\mu}(\lambda)$.

\section{Numerical results}
\label{sec:num}

We are now in a position to present our numerical results.
We first describe our theoretical input and the kinematic conditions.
We use $m_c=M/2=(1.5\pm0.1)$~GeV and the LO formula for
$\alpha_s^{(n_f)}(\mu_r)$ with $n_f=3$ \cite{pdg}.
As for the proton PDFs, we employ the LO set by Martin, Roberts, Stirling, and 
Thorne (MRST98LO) \cite{mrst}, with asymptotic scale parameter
$\Lambda^{(4)}=174$~MeV, as our default and the LO set by the CTEQ
Collaboration (CTEQ5L) \cite{cteq}, with $\Lambda^{(4)}=192$~MeV, for
comparison.
The corresponding values of $\Lambda^{(3)}$ are 204~MeV and 224~MeV, 
respectively.
We choose the renormalization and factorization scales to be $\mu_i=\xi_im_T$,
with $i=r,f$, respectively, and independently vary the scale parameters
$\xi_r$ and $\xi_f$ between 1/2 and 2 about the default value 1.
We adopt the values of
$\left\langle{\cal O}^{\chi_{c0}}\left[{}^3\!P_0^{(1)}\right]\right\rangle$ 
and
$\left\langle{\cal O}^{\chi_{c0}}\left[{}^3\!S_1^{(8)}\right]\right\rangle$
from Ref.~\cite{bkl}.
Specifically, the former was extracted from the measured partial decay widths
of $\chi_{c2}\to\gamma\gamma$ \cite{pdg}, while the latter was fitted to the
$p_T$ distribution of $\chi_{cJ}$ inclusive hadroproduction \cite{abe} and the
cross-section ratio $\sigma_{\chi_{c2}}/\sigma_{\chi_{c1}}$ \cite{aff}
measured at the Tevatron.
For set MRST98LO, these values read
$\left\langle{\cal O}^{\chi_{c0}}\left[{}^3\!P_0^{(1)}\right]\right\rangle
=(8.9\pm1.3)\times10^{-2}$~GeV${}^5$ and
$\left\langle{\cal O}^{\chi_{c0}}\left[{}^3\!S_1^{(8)}\right]\right\rangle
=(2.3\pm0.3)\times10^{-3}$~GeV${}^3$.
The corresponding values for set CTEQ5L are
$(9.1\pm1.3)\times10^{-2}$~GeV${}^5$ and
$(1.9\pm0.2)\times10^{-3}$~GeV${}^3$, respectively.
In order to estimate the theoretical uncertainties in our predictions, we vary
the unphysical parameters $\xi_r$ and $\xi_f$ as indicated above, take into
account the experimental errors on $m_c$ and the default MEs, and switch from
our default PDF set to the CTEQ5L one, properly adjusting $\Lambda^{(3)}$ and
the MEs.
We then combine the individual shifts in quadrature, allowing for the upper
and lower half-errors to be different.

\begin{figure}[ht]
\begin{center}
\epsfysize=18cm
\epsfbox[69 64 520 712]{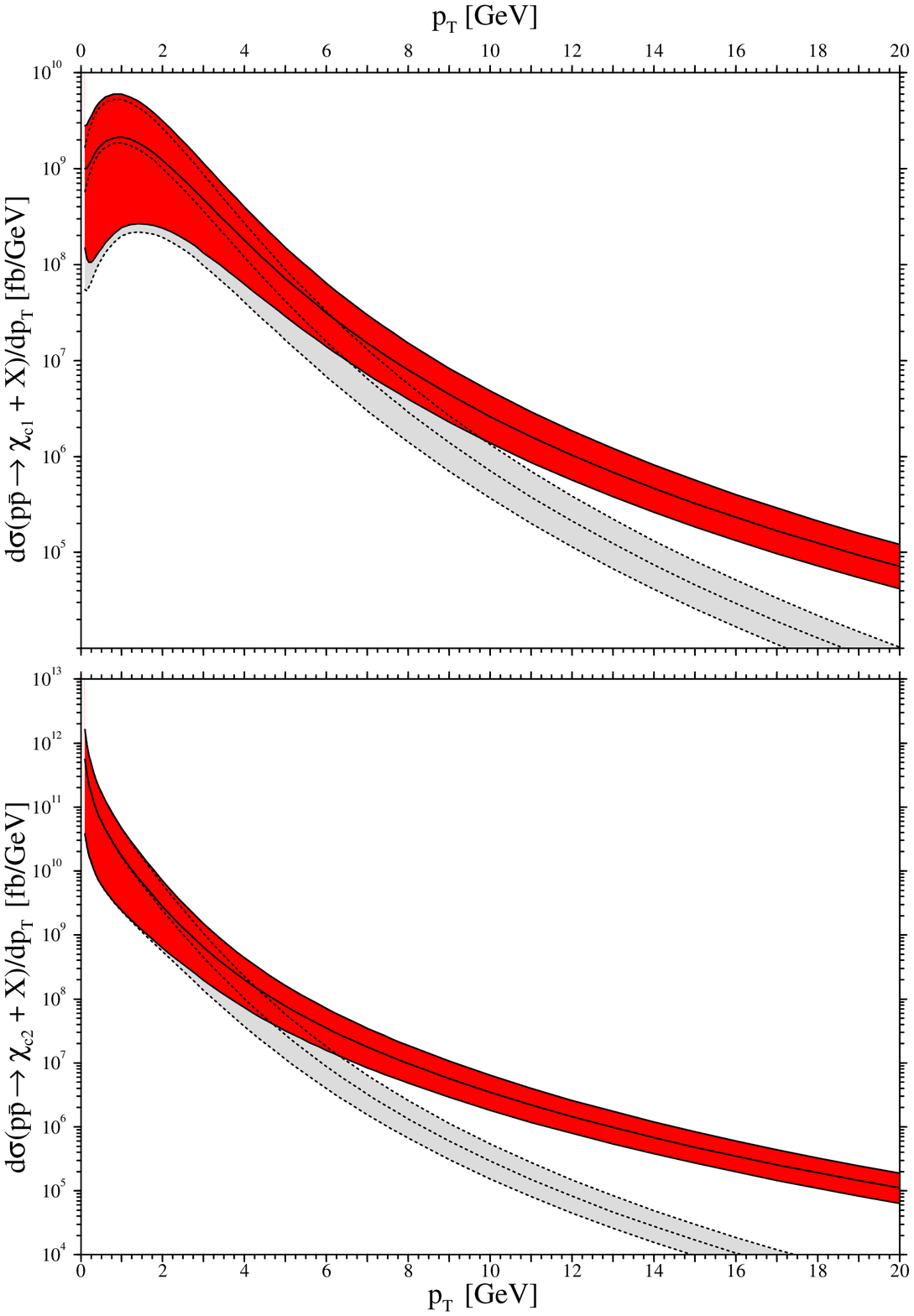}
\caption{Differential cross sections $d\sigma/dp_T$ of
$p\overline{p}\to\chi_{cJ}+X$ for $J=1$ (upper frame) and $J=2$ (lower frame)
at \protect$\sqrt S=1.8$~TeV in NRQCD (solid lines) and the CSM (dashed lines).
The central lines represent the default predictions, and the shaded bands
indicate the theoretical uncertainties.
\label{fig:dsdpt}}
\end{center}
\end{figure}

\begin{figure}[ht]
\begin{center}
\epsfysize=18cm
\epsfbox[55 65 540 760]{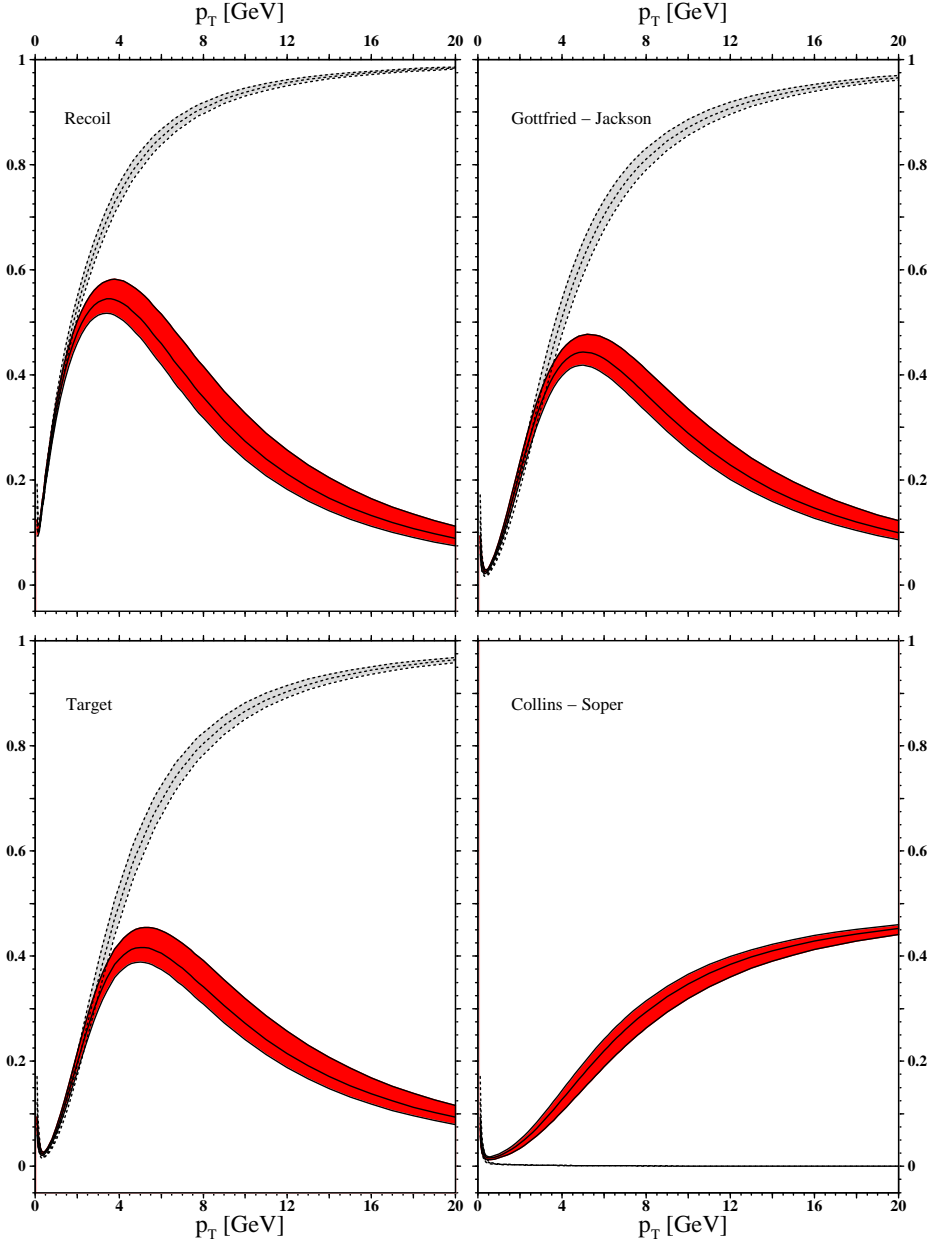}
\caption{Differential helicity density matrix element $d\rho_{0,0}^1/dp_T$ of
$p\overline{p}\to\chi_{cJ}+X$ at \protect$\sqrt S=1.8$~GeV in four different
polarization frames in NRQCD (solid lines) and the CSM (dashed lines).
The central lines represent the default predictions, and the shaded bands
indicate the theoretical uncertainties.
\label{fig:rnn1pt}}
\end{center}
\end{figure}

\begin{figure}[ht]
\begin{center}
\epsfysize=18cm
\epsfbox[55 65 540 760]{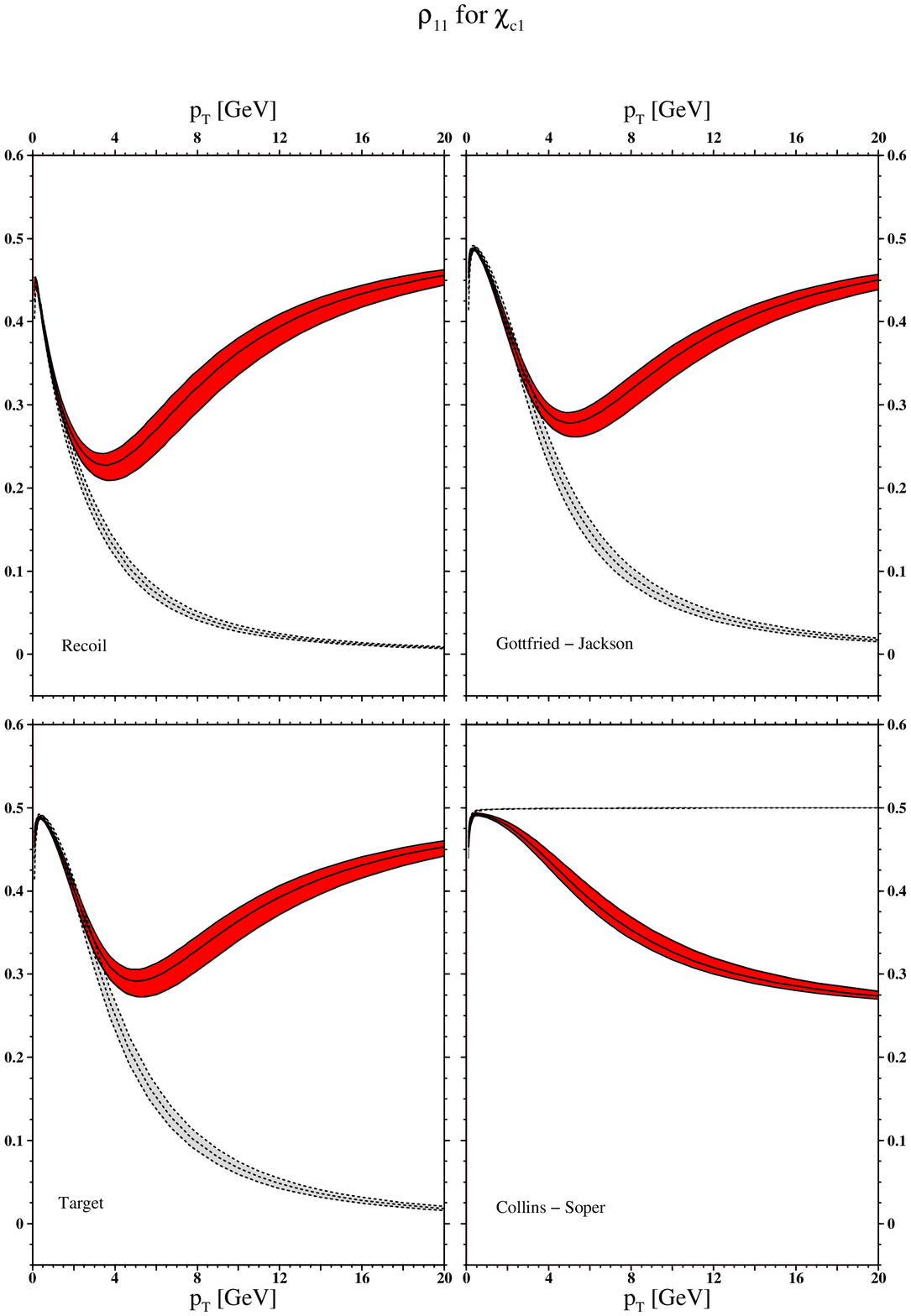}
\caption{Same as in Fig.~\ref{fig:rnn1pt}, but for $d\rho_{1,1}^1/dp_T$.
\label{fig:rpp1pt}}
\end{center}
\end{figure}

\begin{figure}[ht]
\begin{center}
\epsfysize=18cm
\epsfbox[55 65 540 760]{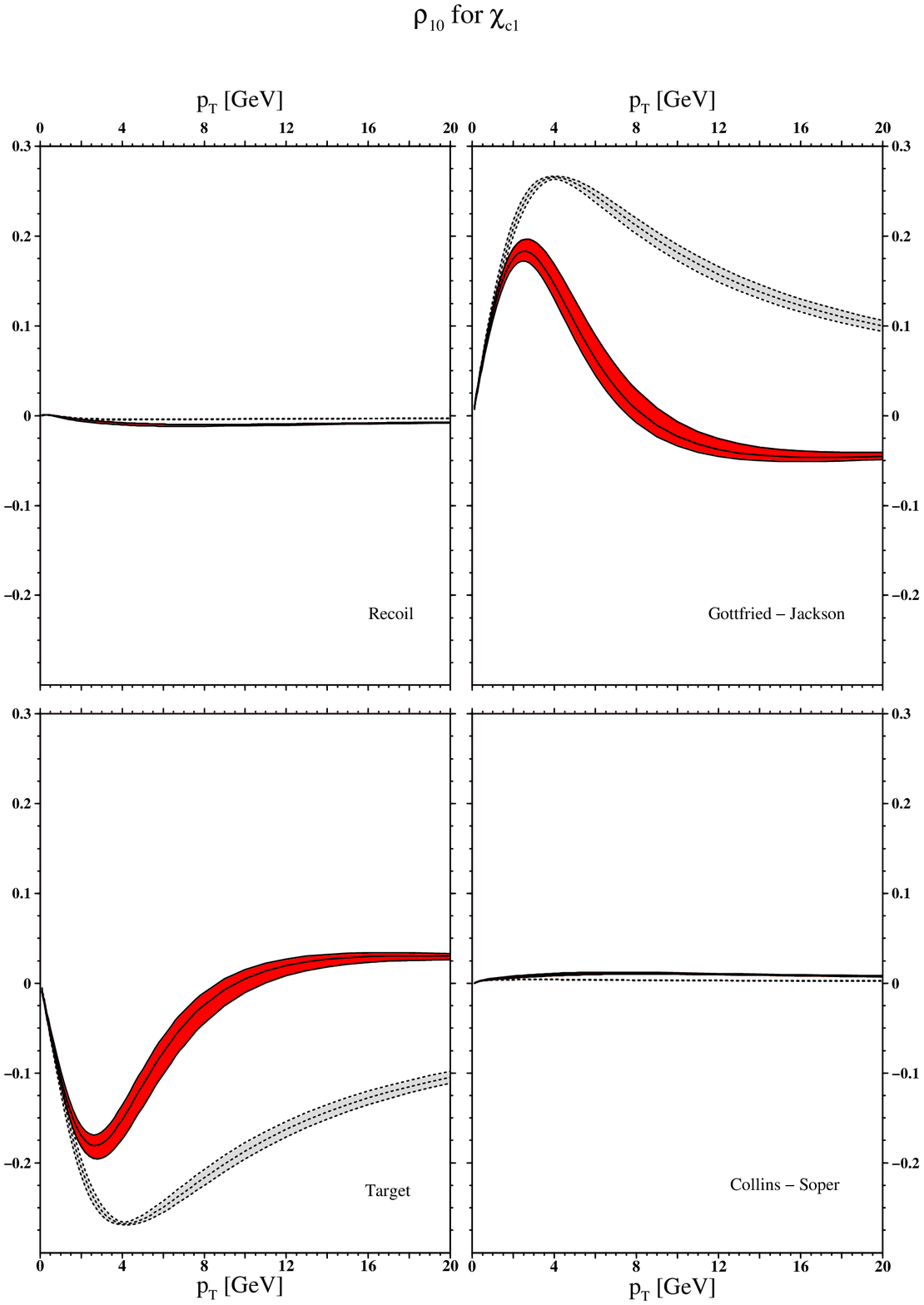}
\caption{Same as in Fig.~\ref{fig:rnn1pt}, but for $d\re\rho_{1,0}^1/dp_T$.
\label{fig:rpn1pt}}
\end{center}
\end{figure}

\begin{figure}[ht]
\begin{center}
\epsfysize=18cm
\epsfbox[55 65 540 760]{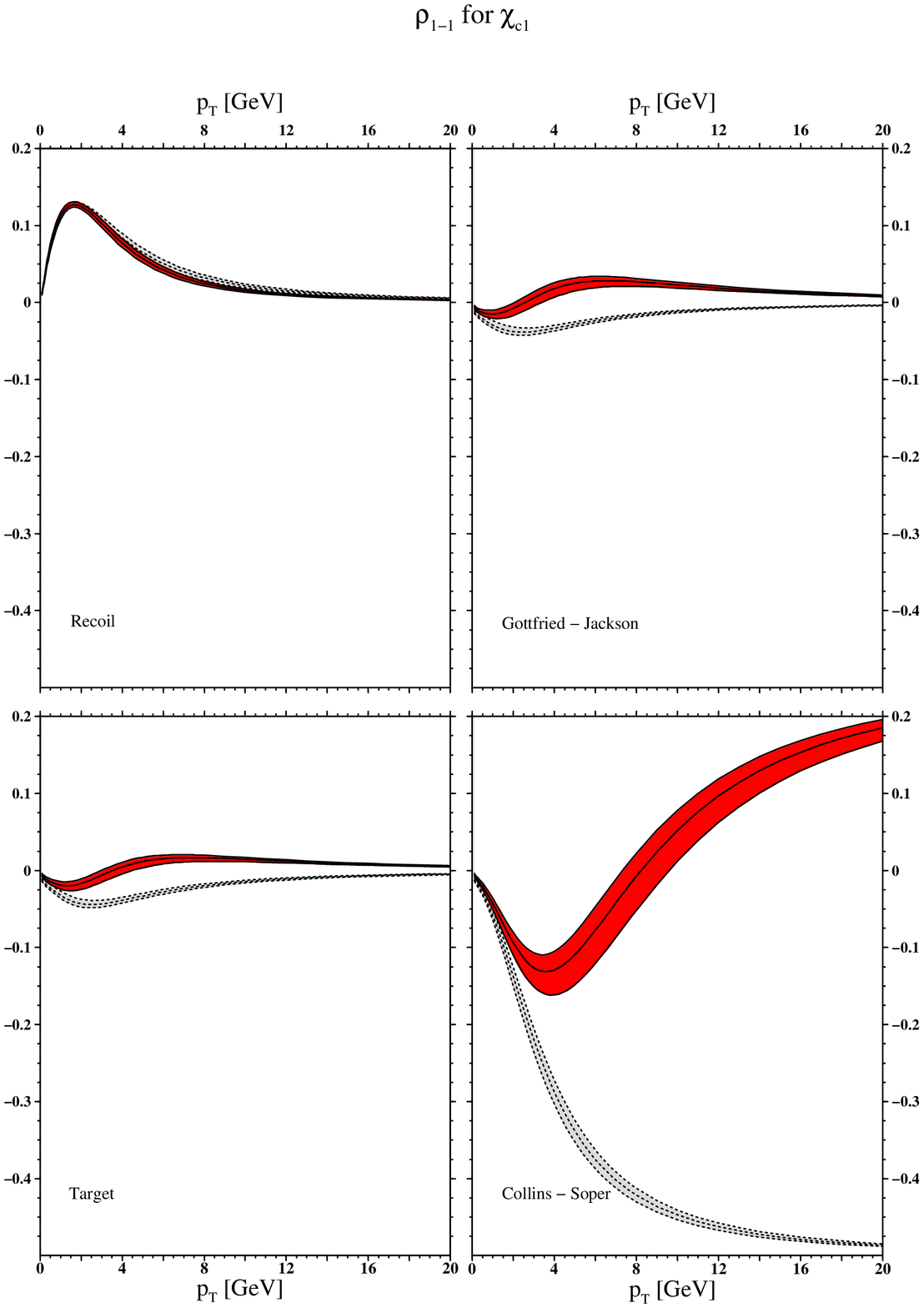}
\caption{Same as in Fig.~\ref{fig:rnn1pt}, but for $d\rho_{1,-1}^1/dp_T$.
\label{fig:rpm1pt}}
\end{center}
\end{figure}

\begin{figure}[ht]
\begin{center}
\epsfysize=18cm
\epsfbox[55 65 540 760]{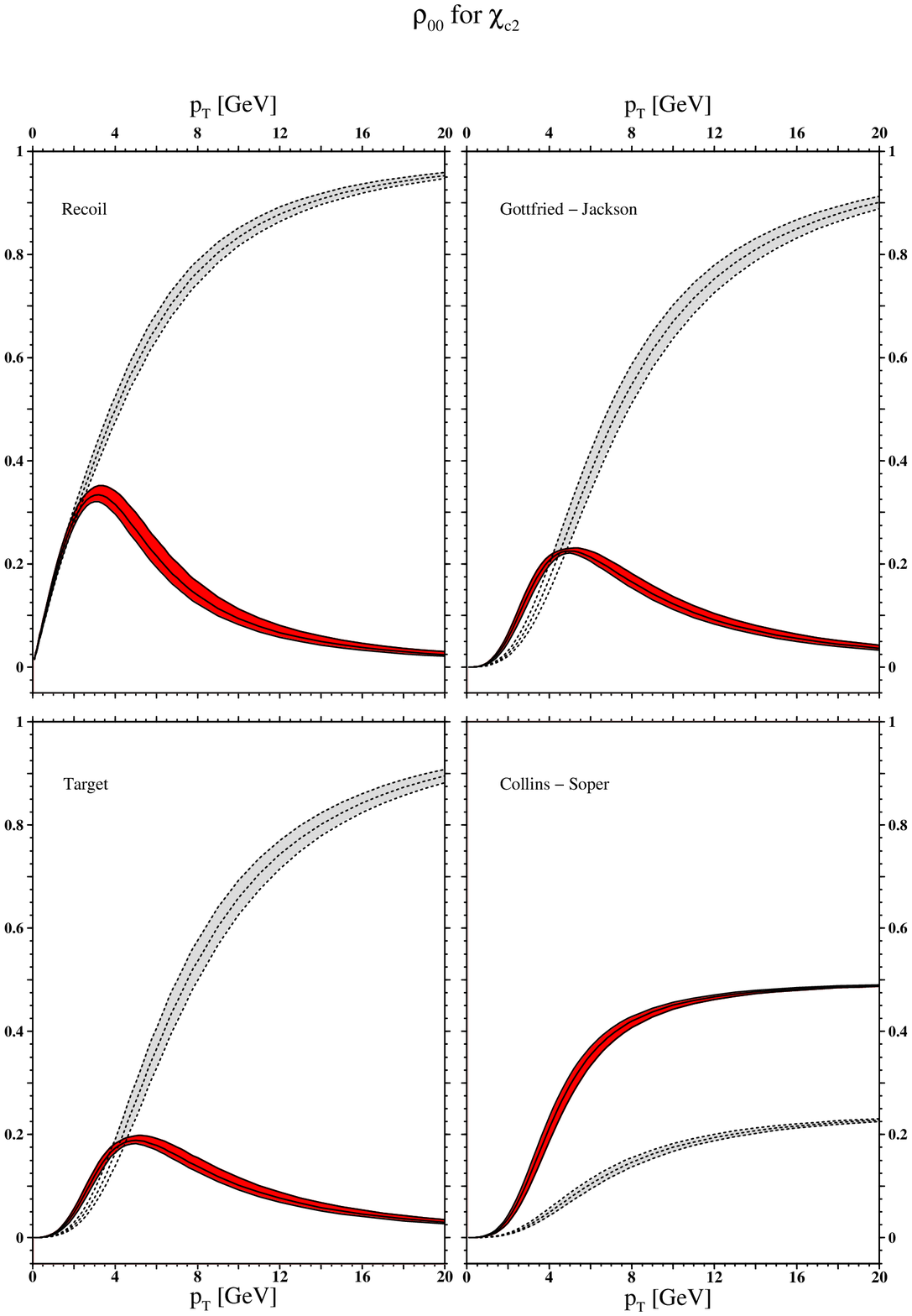}
\caption{Same as in Fig.~\ref{fig:rnn1pt}, but for $d\rho_{0,0}^2/dp_T$.
\label{fig:Rnn2pt}}
\end{center}
\end{figure}

\begin{figure}[ht]
\begin{center}
\epsfysize=18cm
\epsfbox[55 65 540 760]{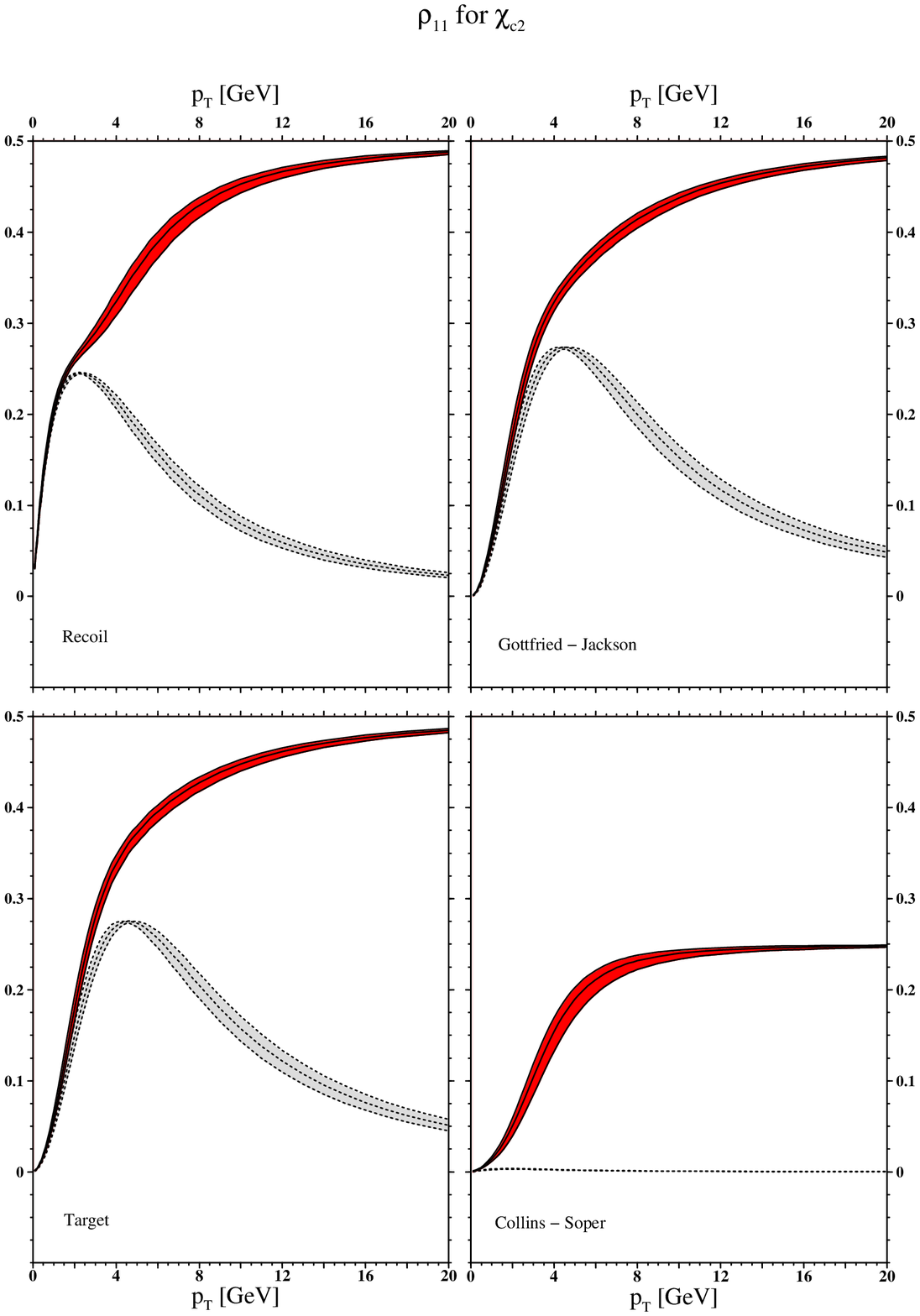}
\caption{Same as in Fig.~\ref{fig:rnn1pt}, but for $d\rho_{1,1}^2/dp_T$.
\label{fig:Rpp2pt}}
\end{center}
\end{figure}

\begin{figure}[ht]
\begin{center}
\epsfysize=18cm
\epsfbox[55 65 540 760]{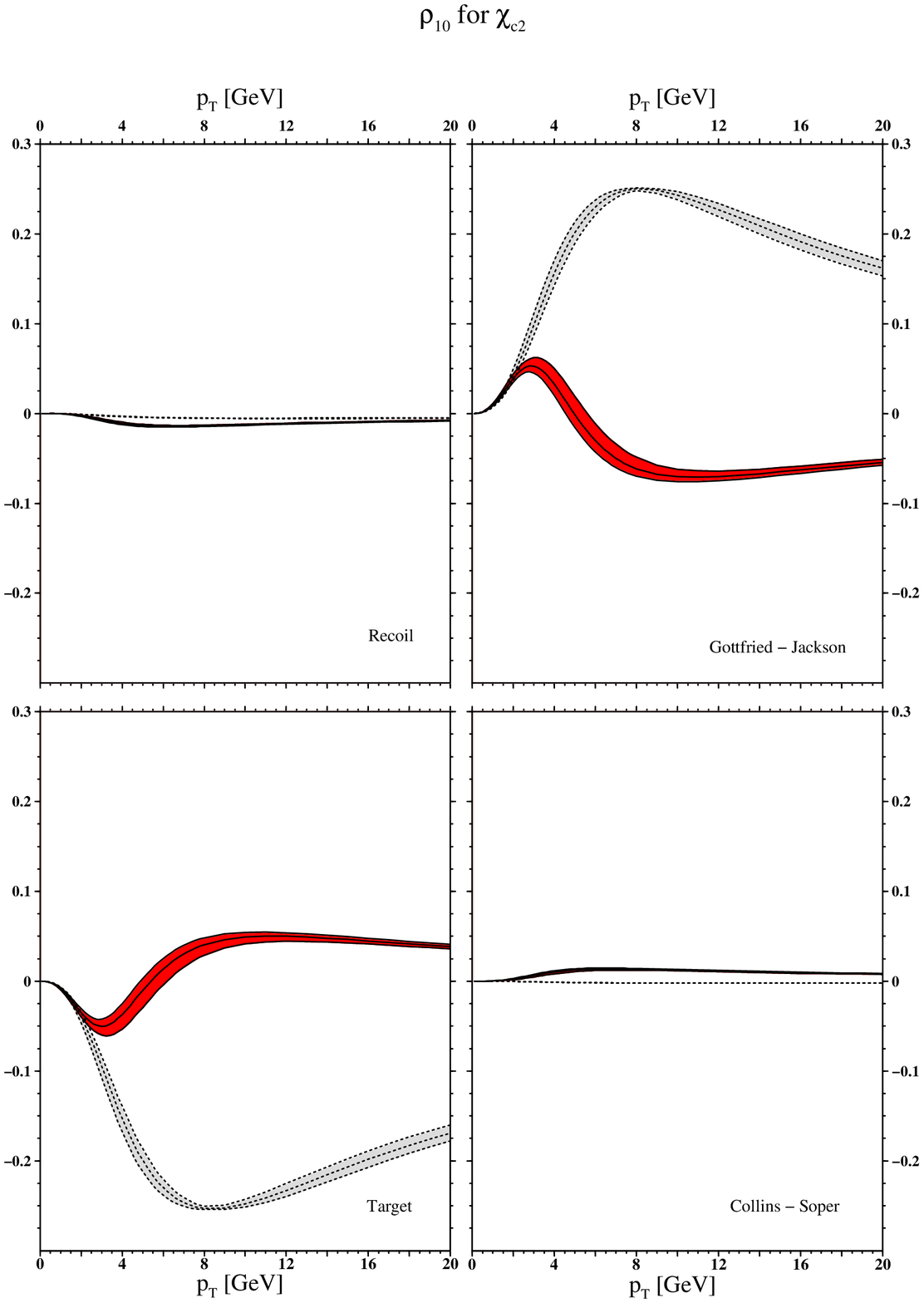}
\caption{Same as in Fig.~\ref{fig:rnn1pt}, but for $d\re\rho_{1,0}^2/dp_T$.
\label{fig:Rpn2pt}}
\end{center}
\end{figure}

\begin{figure}[ht]
\begin{center}
\epsfysize=18cm
\epsfbox[55 65 540 760]{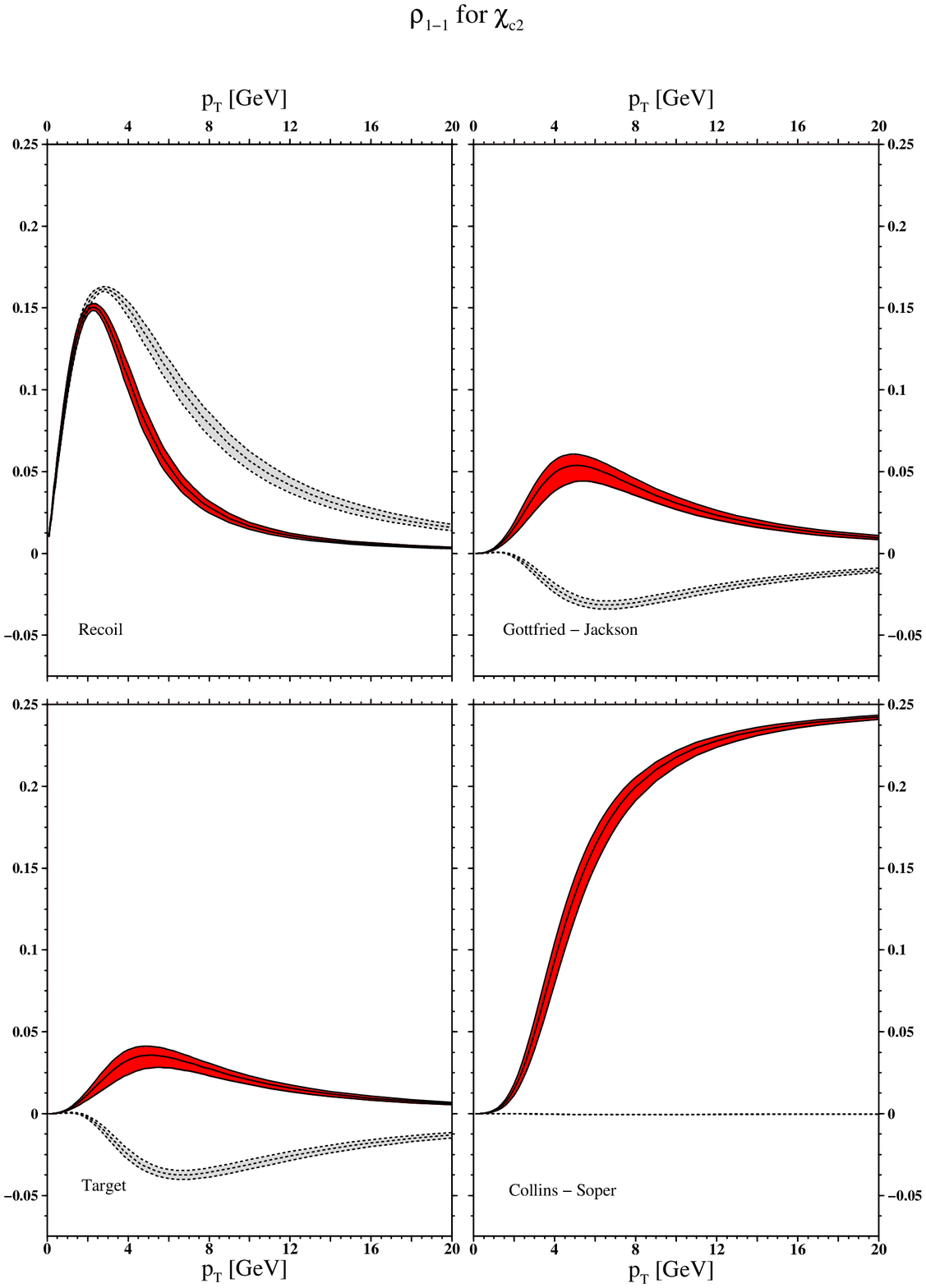}
\caption{Same as in Fig.~\ref{fig:rnn1pt}, but for $d\rho_{1,-1}^2/dp_T$.
\label{fig:Rpm2pt}}
\end{center}
\end{figure}

\begin{figure}[ht]
\begin{center}
\epsfysize=18cm
\epsfbox[55 65 540 760]{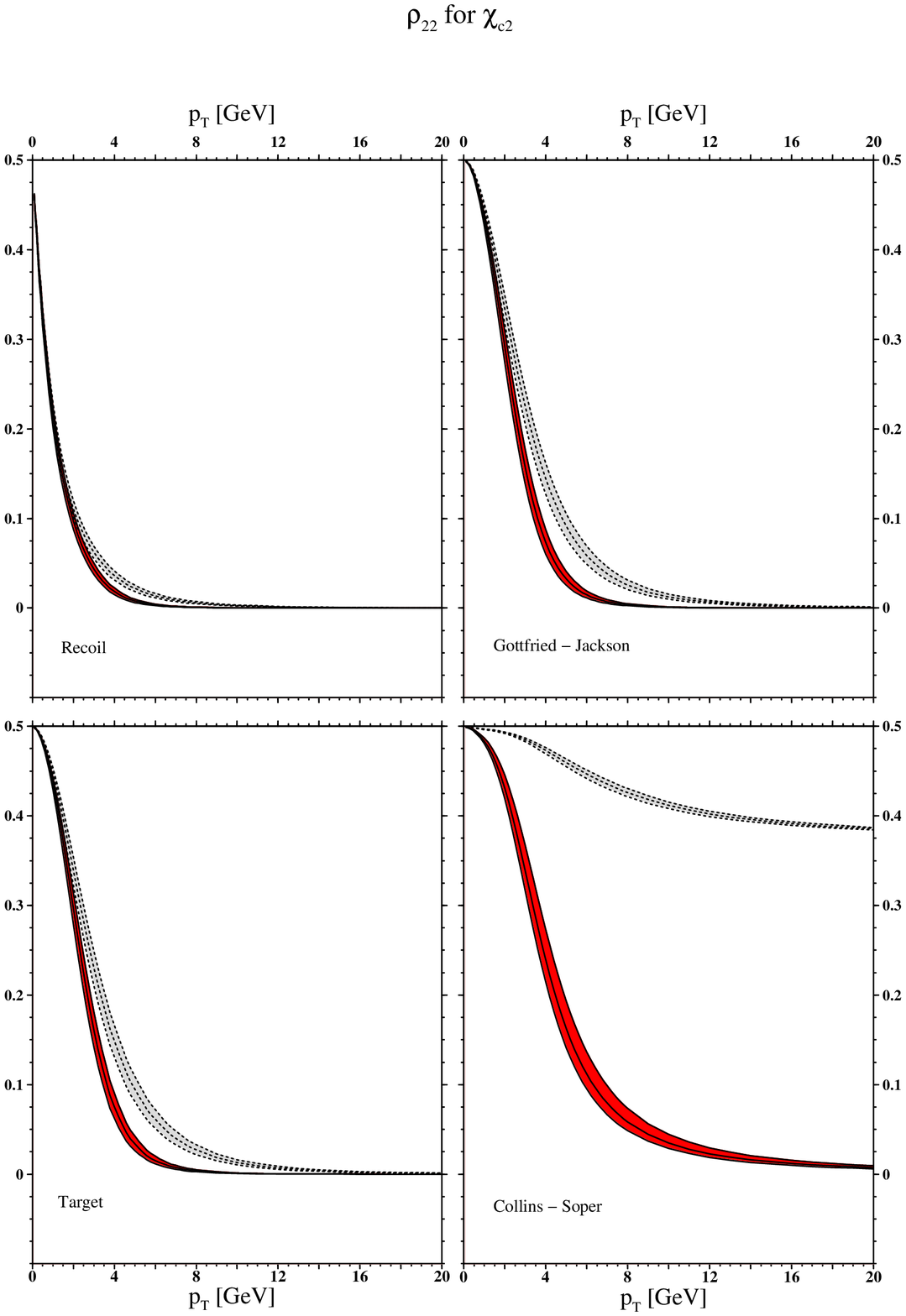}
\caption{Same as in Fig.~\ref{fig:rnn1pt}, but for $d\rho_{2,2}^2/dp_T$.
\label{fig:Rzz2pt}}
\end{center}
\end{figure}

\begin{figure}[ht]
\begin{center}
\epsfysize=18cm
\epsfbox[55 65 540 760]{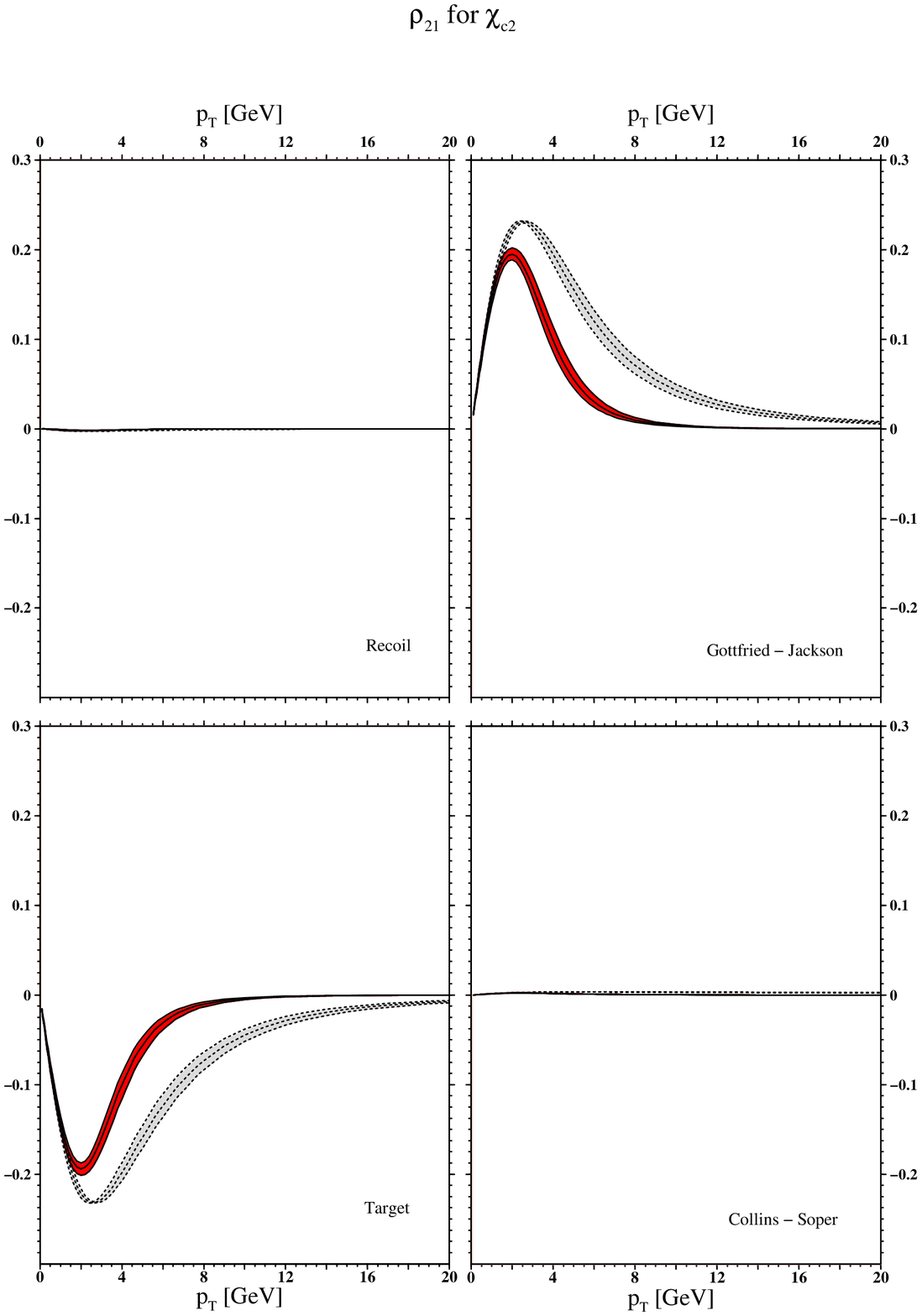}
\caption{Same as in Fig.~\ref{fig:rnn1pt}, but for $d\re\rho_{2,1}^2/dp_T$.
\label{fig:Rzp2pt}}
\end{center}
\end{figure}

\begin{figure}[ht]
\begin{center}
\epsfysize=18cm
\epsfbox[55 65 540 760]{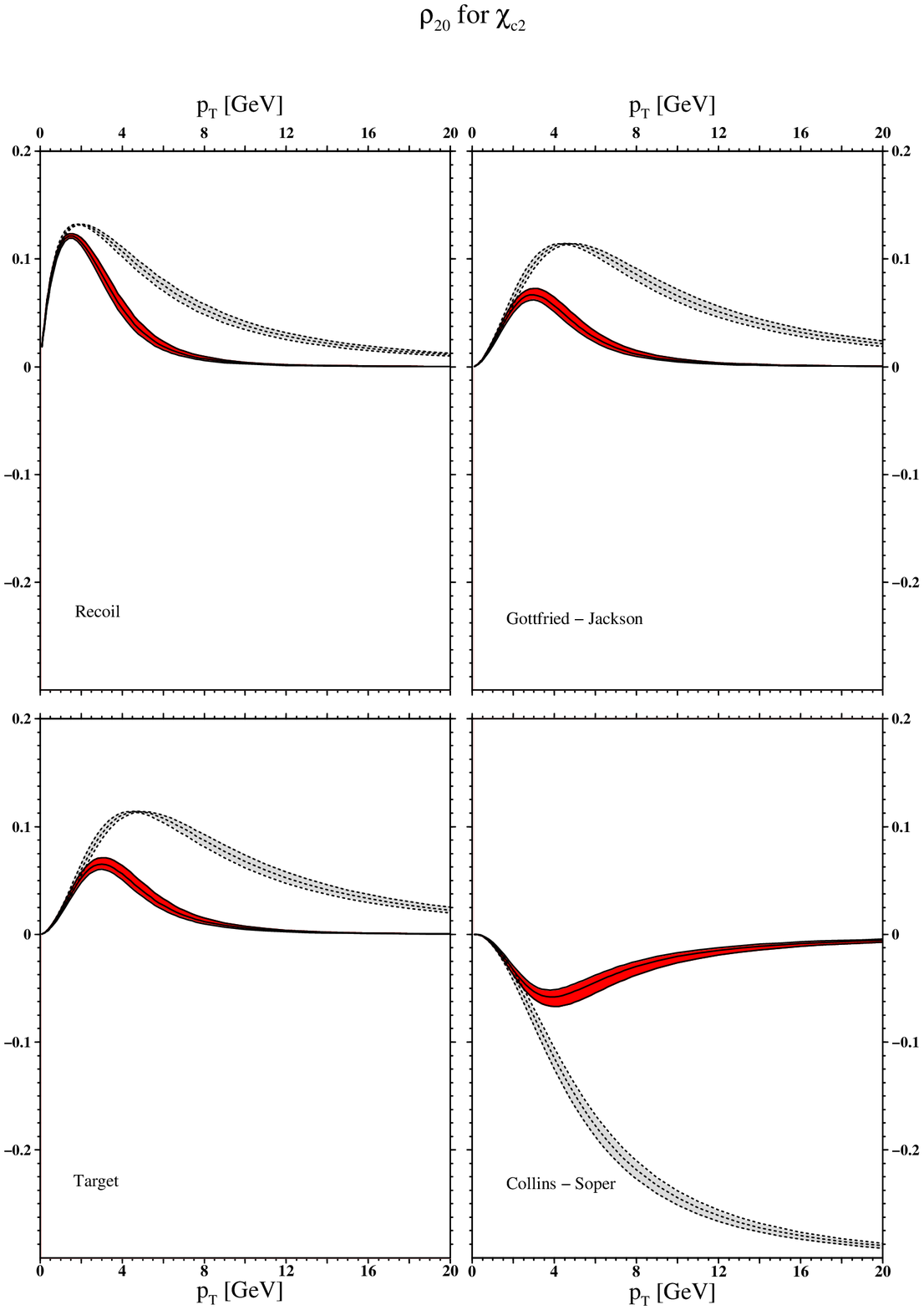}
\caption{Same as in Fig.~\ref{fig:rnn1pt}, but for $d\re\rho_{2,0}^2/dp_T$.
\label{fig:Rzn2pt}}
\end{center}
\end{figure}

\begin{figure}[ht]
\begin{center}
\epsfysize=18cm
\epsfbox[55 65 540 760]{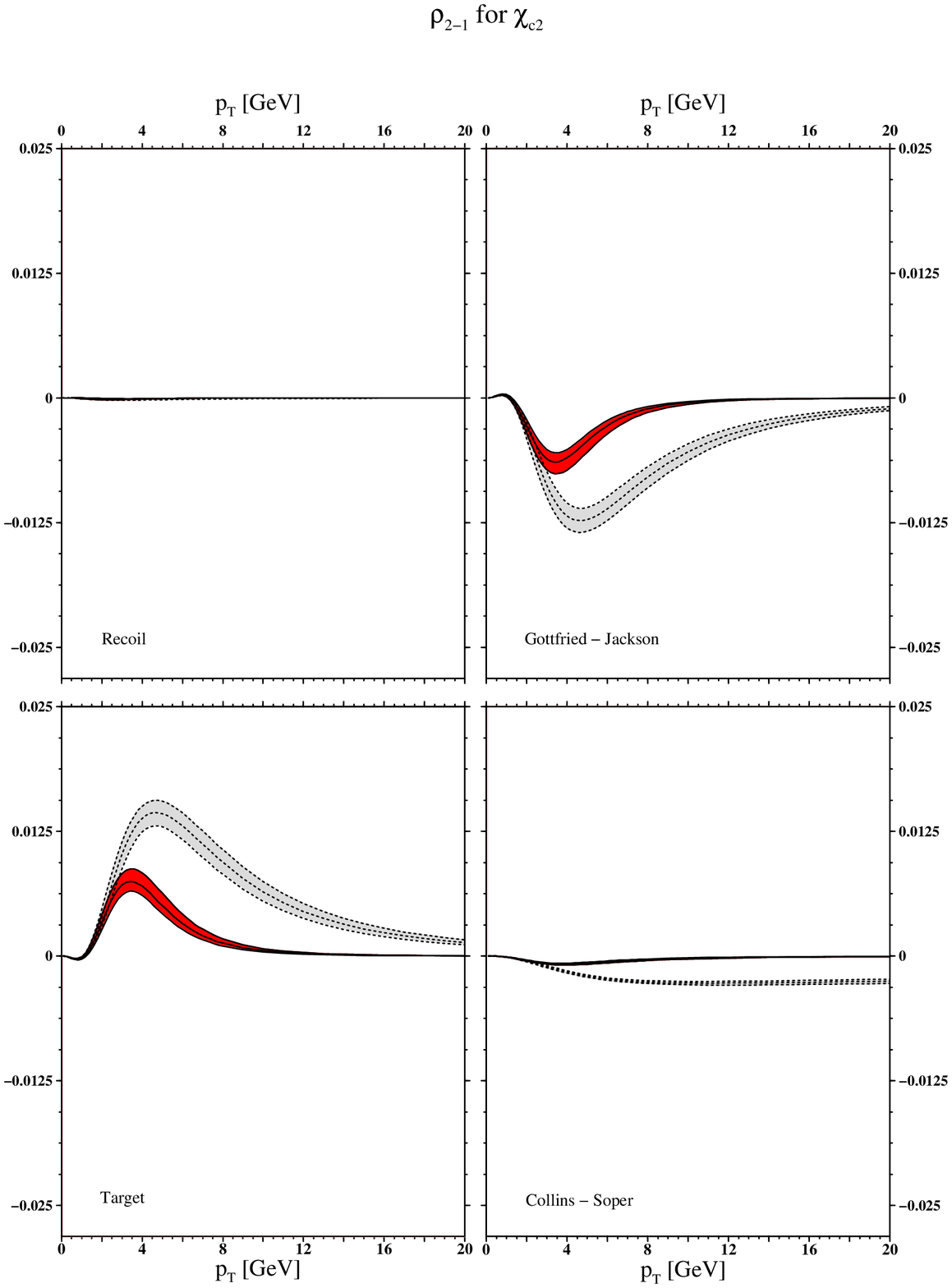}
\caption{Same as in Fig.~\ref{fig:rnn1pt}, but for $d\re\rho_{2,-1}^2/dp_T$.
\label{fig:Rzm2pt}}
\end{center}
\end{figure}

\begin{figure}[ht]
\begin{center}
\epsfysize=18cm
\epsfbox[55 65 540 760]{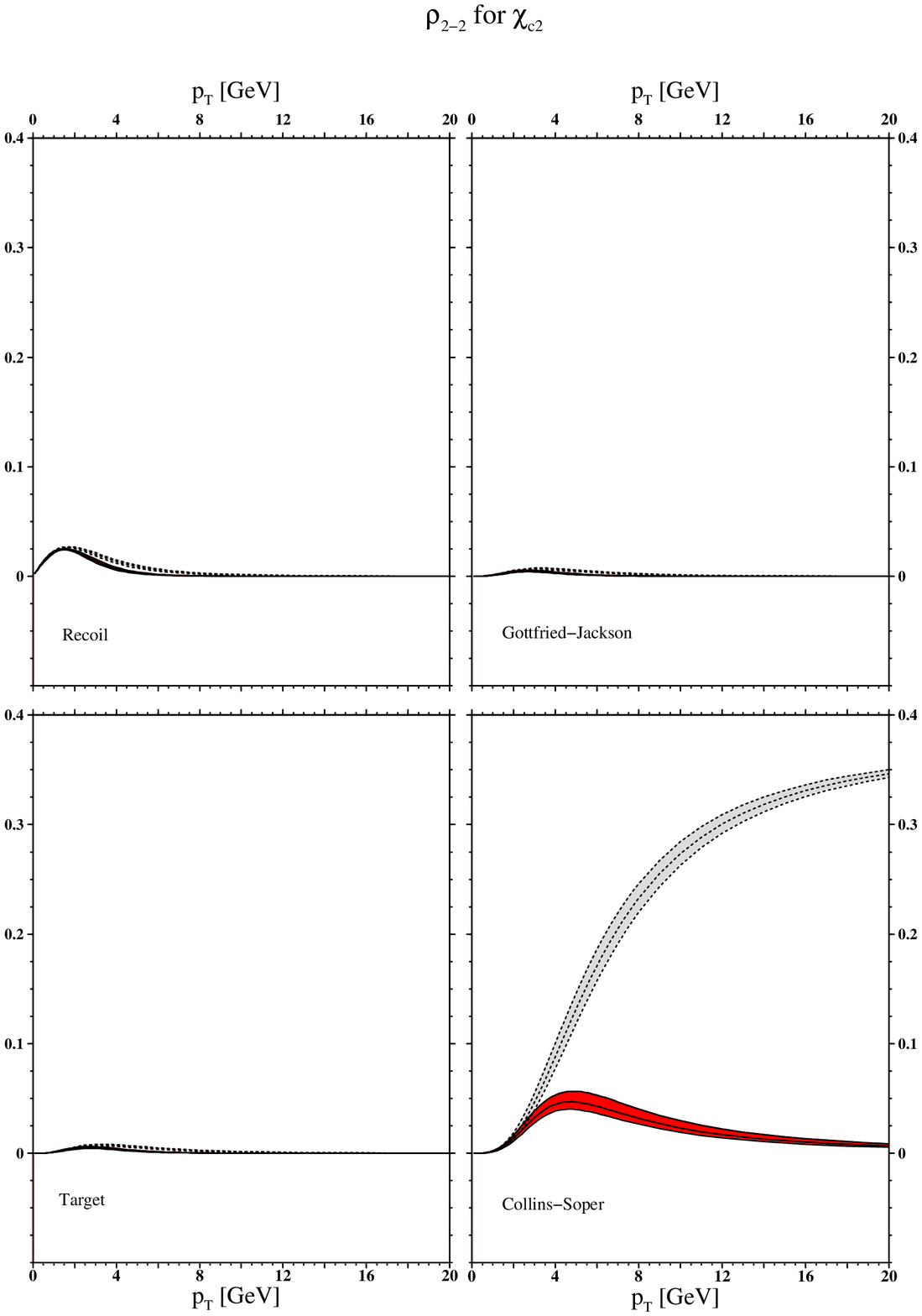}
\caption{Same as in Fig.~\ref{fig:rnn1pt}, but for $d\rho_{2,-2}^2/dp_T$.
\label{fig:Rzi2pt}}
\end{center}
\end{figure}

Our numerical results are presented in
Figs.~\ref{fig:dsdpt}--\ref{fig:Rzi2pt}.
Figure~\ref{fig:dsdpt} is devoted to the cross sections $\sigma$ of
$p\overline{p}\to\chi_{c1}+X$ (upper frame) and $p\overline{p}\to\chi_{c2}+X$
(lower frame), Figs.~\ref{fig:rnn1pt}--\ref{fig:rpm1pt} to the helicity matrix
elements $\rho_{\lambda\lambda^\prime}^J$ of the former process, and
Figs.~\ref{fig:Rnn2pt}--\ref{fig:Rzi2pt} to those of the latter one.
The matrices $\rho_{\lambda\lambda^\prime}^J$ are normalized so that their
traces,
$\sum_{\lambda=-1}^{1}\rho_{\lambda\lambda}^1=2\rho_{11}^1+\rho_{00}^1$ and
$\sum_{\lambda=-2}^{2}\rho_{\lambda\lambda}^2=2\rho_{22}^2+2\rho_{11}^2+
\rho_{00}^2$, are unity.
We only display the real parts of $\rho_{\lambda\lambda^\prime}^J$, which 
enters Eq.~(\ref{eq:w}).
In each figure, the NRQCD (solid lines) and CSM (dashed lines) results are
displayed as functions of $p_T$;
the central lines indicate the default predictions, and the shaded bands the 
theoretical uncertainties.
In Figs.~\ref{fig:rnn1pt}--\ref{fig:Rzi2pt}, four different polarization
frames are considered: the recoil, Gottfried-Jackson, target, and
Collins-Soper frames.
The results in the Gottfried-Jackson and target frames almost coincide if
$|\lambda-\lambda^\prime|$ is even.
If $|\lambda-\lambda^\prime|$ is odd, the same is true, apart from a relative
minus sign.

We first discuss $d\sigma/dp_T$.
From Fig.~\ref{fig:dsdpt}, we observe that the CSM contributions essentially
exhaust the NRQCD results at small values of $p_T$, while they get rapidly
suppressed as the value of $p_T$ increases.
This may be understood by observing that, with increasing value of $p_T$, the
processes (\ref{eq:pro}) with $n={}^3\!S_1^{(8)}$ gain relative importance,
since their cross sections involve a gluon propagator with small virtuality,
$q^2=M^2$, and are, therefore, enhanced by powers of $p_T^2/M^2$ relative to
those of the other contributing processes.
In the fragmentation picture \cite{kra}, these cross sections would be
evaluated by convoluting those of $gg\to gg$, $gq\to gq$, and
$q\overline{q}\to gg$ with the
$g\to c\overline{c}\left[{}^3\!S_1^{(8)}\right]$ fragmentation function
\cite{yua}.
For such processes, the attribute {\it fragmentation prone} has been coined
\cite{pal}.

We now turn to $d\rho_{\lambda\lambda^\prime}^1/dp_T$.
Looking at Figs.~\ref{fig:rnn1pt}--\ref{fig:rpm1pt}, we observe that the NRQCD
and CSM predictions are generally rather similar in the low-$p_T$ range.
However, at large values of $p_T$, there may be dramatic differences,
depending on the matrix element $\rho_{\lambda\lambda^\prime}^1$ considered
and the polarization frame chosen.
Specifically, the diagonal elements, $\rho_{0,0}^1$ and $\rho_{1,1}^1$, lend
themselves as powerful discriminators between NRQCD and CSM in all four
polarization frames.
In the case of $\rho_{1,0}^1$, only the Gottfried-Jackson and target frames
are useful, while the Collins-Soper frame is preferable in connection with
$\rho_{1,-1}^1$.

We finally draw our attention to $d\rho_{\lambda\lambda^\prime}^2/dp_T$ (see
Figs.~\ref{fig:Rnn2pt}--\ref{fig:Rzi2pt}).
Again, the NRQCD and CSM predictions always merge in the limit $p_T\to0$.
On the other hand, in the large-$p_T$ regime, we can always find a 
polarization frame that allows us to clearly distinguish NRQCD from the CSM.
As for $\rho_{0,0}^2$ and $\rho_{1,1}^2$, all four frames can serve this 
purpose.
The Gottfried-Jackson and target frames work best for $\rho_{1,0}^2$,
$\rho_{2,1}^2$, and $\rho_{2,-1}^2$, while the Collins-Soper frame is the 
frame of choice for $\rho_{1,-1}^2$, $\rho_{2,2}^2$, $\rho_{2,0}^2$, and
$\rho_{2,-2}^2$.
We observe that the NRQCD results for $\rho_{1,0}^2$, $\rho_{2,0}^2$,
$\rho_{2,-1}^2$, and $\rho_{2,-2}^2$ are numerically suppressed, with
magnitudes of order 0.1 or below.
Notice that the NRQCD and CSM results for $\rho_{2\lambda^\prime}^2$
($\lambda^\prime=2,1,0,-1,-2$) differ, although the only contributing
$c\overline{c}$ Fock state with $J=2$ is a CS state, $n={}^3\!P_2^{(1)}$.
This may be understood by observing that the CO contribution enters through
the normalization of $\rho_{\lambda\lambda^\prime}^2$.

\section{Conclusions}
\label{sec:con}

Run 2 at the Tevatron, which has just begun, will provide us with a wealth of
new information on how $c\overline{c}$ pairs turn into physical charmonia.
In particular, this will allow us to put the NRQCD factorization hypothesis to
a stringent test.
In the face of this exciting situation, we considered the processes
$p\overline{p}\to\chi_{cJ}+X$ for $J=1,2$ followed by the radiative decays
$\chi_{cJ}\to J/\psi\gamma$ under Tevatron kinematic conditions in the NRQCD
factorization formalism, and we investigated the decay angular distributions,
using the helicity density matrix formalism, with regard to their power to
identify the presence of CO processes.
Specifically, we expressed the distributions in the $J/\psi$ polar and
azimuthal angles in the $\chi_{cJ}$ rest frame in terms of the helicity
density matrix elements $\rho_{\lambda\lambda^\prime}^J$ of the $\chi_{cJ}$
production processes.
We then analyzed the matrix elements $\rho_{\lambda\lambda^\prime}^J$ as
functions of the $\chi_{cJ}$ transverse momentum $p_T$ in four frequently used
polarization frames, namely, the recoil, Gottfried-Jackson, target, and
Collins-Soper frames.
We found that the CS processes play the leading role in the low-$p_T$ range.
This fact could be exploited to extract the CS ME
$\left\langle{\cal O}^{\chi_{c0}}\left[{}^3\!P_0^{(1)}\right]\right\rangle$.
At large values of $p_T$, typically for $p_T\agt5$~GeV, the NRQCD and CSM
predictions can differ significantly, depending on the matrix element
$\rho_{\lambda\lambda^\prime}^J$ considered and the polarization frame
selected.
We could demonstrate that for every matrix element
$\rho_{\lambda\lambda^\prime}^J$ there is at least one polarization frame
where the NRQCD and CSM results are greatly different.
The decay angular distributions under consideration here should, therefore,
lend themselves as sensitive probes of the CO processes to be doubtlessly
established by experiment.

\bigskip

\centerline{\bf Acknowledgments}

\smallskip

We thank Hee Sung Song for raising our interest in this study.
The research of B.A.K. and G.K. was supported in part by the DFG through Grant
No.\ KN~365/1 and by the BMBF through Grant No.\ 05~HT1GUA/4.
The research of C.P.P. was supported in part by the DFG through
Graduiertenkolleg No.\ GRK~602/1 and by the Office of the Vice President for
Academic Affairs of the University of the Philippines.
This research was performed in the framework of the DFG-KOSEF cooperation
project No.\ 446~KOR-113/137.

\newpage

\newpage

\newpage

\newpage

\newpage

\newpage

\newpage

\newpage

\newpage

\newpage

\newpage

\newpage

\newpage

\newpage

\end{document}